# Geodiversity of Research: A Comparison of Geographical Topic Focus and Author Location using SDG 2: Zero Hunger as a Case Study

Philip J. Purnell

Centre for Science and Technology Studies,

Leiden University,

P.O. Box 905, 2300 AX Leiden, The Netherlands

Tel: +971 50 552 9356

p.j.purnell@cwts.leidenuniv.nl

ORCID: 0000-0003-3146-2737

## Abstract

This study examined the geodiversity of research through comparing topic focus with author location using SDG 2: Zero hunger as a case study. As the research was related to hunger, papers were mapped on to the Global Hunger Index country categories as convenient classification. The publication dataset comprised 60,000 papers from the Dimensions database that have been associated with hunger research using Digital Science's machine learning algorithm that enhances expert led search strategies. Only 41% of hunger-related publications that focus on countries most affected by hunger feature authors affiliated to institutions in those countries. Even fewer of those publications feature locally based authors in first or last position. These numbers gradually reverse as the level of hunger declines. We analyse sample papers in an attempt to understand the reasons for these trends. These included differences in research infrastructure, sub-authorship recognition such as acknowledgements, and limitations of the relationship between country mention and real topical focus. We did not find evidence of widespread differences between senior and overall authorship and consequently urge caution before judging international collaborations as 'helicopter' research based only on author country affiliations and authorship position.

## Keywords

# Introduction

Sustainable development has been a global agenda since the 1980s with increasing involvement of the scientific community (Brundtland, 1987; Hassan, 2001; IUCN–UNEP–WWF, 1980; Kates et al., 2001). During many of these discussions, it has been established that studies of society's most pressing challenges offer more value when conducted at locations where most impact is felt (Balvanera et al., 2017; Mirtl et al., 2018; Sutterlüty et al., 2018). Using bibliometric data, we can conduct large scale studies of the geographical focus of research articles by looking at the titles and abstracts of publications. This macro level analysis enables us to build a picture of which countries researchers are focusing on.

A related question is where researchers are physically located. There is a notable difference between the global, often theoretical perspectives of well-funded academics in countries with established research infrastructure, and the action-based urgency of resource-poor local scientific communities in developing countries (Kates, 2011). There is growing concern about practices of researchers from wealthy nations using samples or data from developing regions (e.g., Amugune & Otieno-Omutoko, 2019; Bockarie, Machingaidze, Nyirenda, Olesen, & Makanga, 2018; van Groenigen & Stoof, 2020). By using author affiliations on published research, we can contribute quantitative analyses to support this discussion.

Cooperation between international scientists and academic communities in developing regions is expected to be conducted in an ethical manner. Guidelines initially centred around the actions of individual scientists (Yakubu et al., 2018) and have recently expanded to call for oversight of how these partnerships are managed by funders, societies, and academic publishers (e.g., Aramesh, 2019; Heinz, Holt, Meehan, & Sheppard, 2021; Nature, 2022; PLOS, 2021). The 2022 World Conference on Research Integrity hosted a discussion that will lead to a "Cape Town Statement" on equitable research partnerships (Horn et al., 2022). Analysis of international collaboration and author position may provide data to further this discussion.

In our study, we use bibliometric data to find out to what extent research is conducted where its impact is most needed. With nearly one in ten people today suffering from chronic hunger (von Grebmer et al., 2020), we chose SDG 2 as the focus of this case study. We used the Dimensions database because it has broader coverage than databases such as Scopus or Web of Science (Hook, Porter, & Herzog, 2018) and indexes a substantial amount of short scientific documents such as meeting abstracts and scientific communications not included in Scopus (Visser, van Eck, & Waltman, 2021). In our view, additional coverage beyond traditional peer-reviewed journals is important in the context of research on local issues in countries with less developed academic publishing systems. Other studies have shown advantages of the Dimensions coverage in the social sciences (van Leeuwen, Engels, & Kulczycki, 2022) and infectious diseases (Rahim, Khakimova, Ebrahimi, Zolotarev, & Rafiei Nasab, 2021). Using Dimensions is therefore a deliberate decision aimed at maximising coverage of relevant scholarly content.

### Research questions

To understand the geodiversity of research into alleviating hunger, we seek to compare the topical focus of scholarly research with individual countries. There are nearly 60,000 research papers related to SDG 2 – Zero Hunger published between 2016 and 2021 and indexed in the Dimensions database. We search within the title and abstract of these papers for mentions of countries as a proxy for their geographical topic focus. We then classify the papers by the severity of hunger in the country of focus using the groups defined by the 2021 GHI. Therefore, our first research question is: *To what extent is research on hunger focused on countries affected by hunger?*

The geographical location of published authors can be established through analysis of the author address or affiliation in each paper. The address usually contains the name of an institution and a country name. We look at the country affiliation of authors in the SDG 2 related research papers and match them with the hunger severity groupings described by the GHI. Our second research question is: *To what extent are scholars involved in research on hunger located in countries severely affected by hunger?*

Research partnerships that aspire to fairness and equity may offer benefits to communities in developing regions (Fransman et al., 2021). Therefore, ethical cooperation between academics in countries afflicted by severe hunger and scholars from wealthy countries with long established research infrastructures is potentially beneficial. Recent editorials have raised concerns about the inequality of international partnerships and even questioned the motives of collaborating researchers from wealthy countries. Analysis of authorship position on the SDG 2 related papers enables us to shed light on the inferred relative seniority of academics from hunger-stricken countries and collaborators from less affected areas. Our third research question is therefore: *To what extent are research publication partnerships between international experts and local scholars equal in research on hunger?*

## Literature review

Mentioning a country name in the title or abstract of an article may indicate topical or localised focus of the study. However, declaring geographical focus may also be associated with socio-cultural patterns related to characteristics of the topic, author location, or sample used in the study (Kahalon, Klein, Ksenofontov, Ullrich, & Wright, 2021). Kahalon et al. (2021) found psychology articles are less likely to mention the country name in an article title if the study includes samples from WEIRD (Western, Educated, Industrialised, Rich, and Democratic) countries. The authors point out that while naming countries in non-WEIRD countries can initially seem inclusive, the practice could be counterproductive by inadvertently suggesting the findings are geography-specific and not generalizable. This suggestion may reinforce the implicit belief that knowledge produced by scientists in and about individuals located in WEIRD countries represent the universal or default position (Castro Torres & Alburez-Gutierrez, 2022). The implication continues that when authors from non-WEIRD countries declare regional focus of their study, these signify exceptions to the rule thereby reducing the articles' usefulness in global research (Kahalon et al., 2021).

The notion of local academics providing adding value to research and knowledge exchange on topics that affect their community is based on deep understanding of the problem and motivation to solve it. The true benefit of the locally grounded research model is achieved by attracting ideas and people into a specific place where they gain access to local knowledge (Billick & Price, 2010; Gerlak et al., 2018; The British Academy, 2021). Reyes-García et al. (2019) note that locally grounded data sources such as indigenous and local knowledge has the potential to contribute an additional layer of knowledge to research on the impact of climate change on social-ecological systems.

Previous studies have lamented the lack of geodiversity among authors in many fields. Conservation science lacks authors from tropical countries that have the most knowledge to contribute and the most impact to suffer from (Mammides et al., 2016). In medical fields, there is a clear under representation of authors from non-high income countries (Campbell et al., 2023; Mooldijk, Licher, & Wolters, 2021). Meanwhile, there may be inherent sampling bias in even the most inclusive global bibliometric databases as demonstrated by the relative lack of linguistic studies on global South languages and by authors based in the global South (Bylund, Khafif, & Berghoff, 2023).

International collaboration including capacity building and joint research projects can play a role in the response to localised social challenges and even humanitarian tragedies (Bajoria, 2011). The 2013-2016 Ebola virus disease outbreak in West Africa generated international attention and stimulated collaboration that involved collaborative clinical studies, training of local outbreak responders, and the establishment of diagnostic and surveillance laboratories (Arias et al., 2016; Heymann, Liu, & Lillywhite, 2016; Yozwiak et al., 2016). Attempts to improve capacity of West African communities to respond to disease outbreaks have been termed 'rooted' collaboration (Yozwiak et al., 2016) because of the co-creation of sustainable local capacity building. Lamentably, not all the international collaborators contributed in the same spirit and some were described as sub-optimal 'parachute' research (Heymann et al., 2016).

Helicopter or parachute research (North, Hastie, & Hoyer, 2020) portrays the image of privileged academics who come to troubled regions from wealthy countries and avoid real collaboration with local scientists. These researchers are said to drop in, collect samples, and leave, sometimes without the knowledge or permission of authorities in the visited country (Heymann et al., 2016). Others have used the term 'colonial research' or 'neo-colonial research' (Minasny & Fiantis, 2018) to suggest that scholars from developed economies feel entitled to take samples from less well developed areas for their own purposes and limit the input of local colleagues.

Where local scientists are excluded from collaboration, or their role is limited to locating and collecting samples, international research teams miss an opportunity to increase the capacity of scholars at the collection site and offer little benefit to the local community (Minasny & Fiantis, 2018). Following the particularly devastating Indonesian peat fires of 2015 that destroyed vast areas of peatland, local scientists set out practical solutions as a priority, and encouraged research into responsible and effective peatland management (Sabiham, Setiawan, Minasny, & Fiantis, 2018). These authors lamented international teams in Indonesia concentrating efforts on more academic research into deforestation, greenhouse gas emissions, and peat fires with less practical value.

As research becomes more collaborative, the number of authors per paper has grown more than five-fold in the last 100 years (Aboukhalil, 2014). That makes the relative contribution of authors less obvious, and readers have turned to proxies such as author position to infer leadership of research projects. In academic literature, first and last author positions are considered 'key' contributors (Mattsson, Sundberg, & Laget, 2011; Wren et al., 2007). The corresponding author also holds considerable weight and has been shown to coincide most frequently with first and then last author position (Mattsson et al., 2011).

Africa-based authors are traditionally underrepresented in the scientific literature. An analysis of author affiliation position on 1,182 biomedical studies conducted in Africa, showed over 93% featured at least one Africa-based co-author (Mbaye et al., 2019). However, Africa-based co-authors featured in fewer than half the articles in first author position, and even fewer in the prestigious last author position. A similar study concurred (Hedt-Gauthier et al., 2019) and called on the research community from high income countries to challenge the established power balance. Despite increased inclusion of local scholars in Africa-based research, inequity remains regarding their relative roles in the team.

The scholarly community has begun to call for action to discourage helicopter or parachute research and instead promote ethical and sustainable collaboration with academics from low- and middle-income countries. A group of editors and researchers published a consensus statement to promote equitable authorship which includes practical advice to journal editors on evaluating manuscript submissions resulting from international collaborations (Morton et al., 2022).

A related discussion has been published in Geoderma, about the prevalence of helicopter research in soil science, along with the limited involvement of local knowledge owners, and the lack of structural improvement in the community at the place of the study (van Groenigen & Stoof, 2020). One participant from Ethiopia suggested building local capacity in young scientists though involvement of students from Masters and PhD programmes who should be first authors where they conducted the work (Haile, 2020). As part of the same discussion, Giller (2020) described how the Dutch research funding agency, NWO-WOTRO Science for Global Development made its grants subject to a compulsory workshop with stakeholders from the country of the proposed study. The workshops often led to closer collaboration between the local and visiting scientists (Giller, 2020). Other journals have taken similar steps to tackle helicopter research and ethics dumping (Morton et al., 2022; Nature, 2022).

Hunger is a complex problem that is interlinked with political or military instability, e.g. South Sudan (Mayai, 2020), Yemen (De Souza, 2017). In regions under civil and military conflict, severe malnutrition contributes to mortality (Leaning & Guha-Sapir, 2013; Salama, Spiegel, Talley, & Waldman, 2004). Under such circumstances, the local education and research infrastructure is often limited (Lai & Thyne, 2007) and much of the work is conducted by researchers in other countries or foreign non-governmental organizations (Ford, Mills, Zachariah, & Upshur, 2009; Kalleberg, 2009). Following high profile events such as famine or war, it is natural to want to help. However in some cases, Western researchers have arrived in affected regions with pre-conceived research protocols thereby limiting the role of local participants and scholars (Asiamah, Awal, & MacLean, 2021). In the context of de-colonising research paradigms, a progressive approach is preferred which involves local scholars and participants in an inductive and iterative way (Firchow & Gellman, 2021; Yom, 2014). Studies of the dynamics between the researcher and the researched have demonstrated benefits in an inclusive approach where those with local knowledge are included at every stage of the research design (Riley, Schouten, & Cahill, 2003) including the very question being addressed. That way, international research teams ought to engage local participants as actors with agency (Gellman, 2021) and avoid limiting their roles.

## Data and Methods

We extracted publications from a version of the Dimensions database hosted by the Centre for Science and Technology Studies (CWTS) at Leiden University. Dimensions uses a multistep process to tag publications considered relevant to sustainable development goals that utilises a machine learning algorithm to enhance expert driven search strategies (Wastl, Fane, Draux, & Diwersy, 2021; Wastl, Hook, Fane, Draux, & Porter, 2020). We selected all records linked to SDG 2: Zero Hunger. The time window used was six full publication years (2016 – 2021). We included journal articles, conference papers, books, monographs, and book chapters, but excluded preprints because they have not been peer-reviewed. We also excluded papers that did not list any author affiliations because the affiliations formed an important part of our analysis.

For country population, we used the most recent UN estimates (The World Bank, 2022).

We counted the number of publications in which a country was mentioned in either the title or abstract of the article. These were termed 'country mentions'. If one or more countries within one of the GHI categories were mentioned in a publication, then it counted as one mention for the category. If one or more countries were mentioned in the publication for two categories, then it counted as one mention for each category.

To determine the geographical location of researchers, we examined the countries in the author affiliations of each paper. These were termed 'country affiliations'. Similar to the mentions, if one or more country affiliation in any one category appeared in a publication, then it counted as one country affiliation for that category. Meanwhile, if one or more country affiliations were found from two GHI categories, it counted as one country affiliation for each category.

For ease of comparison, we present the results in groups of countries, rather than individual nations. As the case study is on hunger related research, we classified papers by country according to the categories described in the Global Hunger Index 2021 report (von Grebmer et al., 2021) and shown in Figure 1 with full country listing in Appendix A.

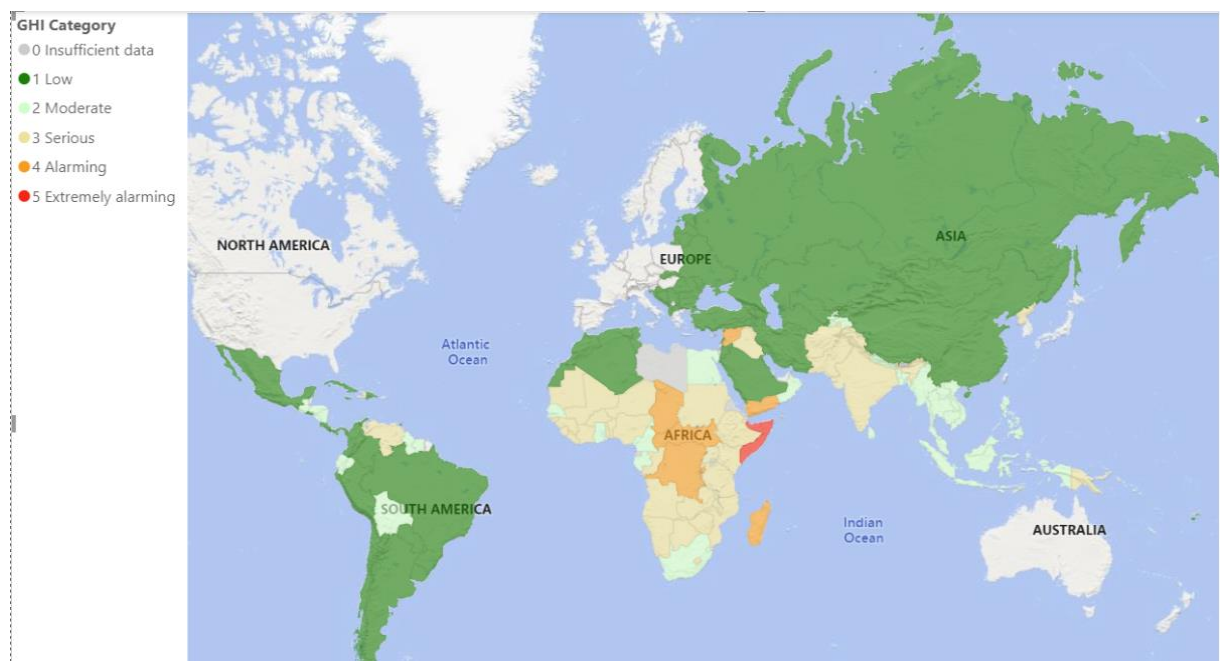

Figure 1. GHI country categories based on the severity of hunger.

In order to create these categories, the GHI ranked countries using a composite score based on four indicators (Table 1); undernourishment, child wasting, child stunting, and child mortality (Wiesmann, Biesalski, von Grebmer, & Bernstein, 2015). There was sufficient data to calculate individual scores for 116 countries. An additional 12 countries were provisionally designated and for seven countries there was insufficient data for even a provisional categorisation. We assigned countries to categories if they were assigned or provisionally assigned by the GHI 2021 report but did not include the remaining seven countries in the study.

Table 1. Global Hunger Index composition.

| Dimension | Indicator | Weighting | Source |
| --- | --- | --- | --- |
| Inadequate food supply | Calorie deficiency Proportion of the population that is undernourished | 1/3 | UN Food and Agriculture Organization (FAO) |
| Child undernutrition | Proportion of children <5 yrs. suffering from stunting | 1/6 | UN United Nations Children's Fund (UNICEF) World Health Organization (WHO) World Bank |
| Child undernutrition | Proportion of children <5 yrs. suffering from wasting | 1/6 | UN United Nations Children's Fund (UNICEF) World Health Organization (WHO) World Bank |
| Child mortality | Child <5 yrs. mortality rate | 1/3 | UN Inter-Agency Group for Child Mortality Estimation (UN IGME) |

Source: Global Hunger Index 2021 (von Grebmer et al., 2021)

The International Food Policy Research Institute (IFPRI) was founded in 1975 and organises research projects in the areas of food supply, nutrition, food trade systems, agricultural economies, and governance. It also runs country level research programmes because of the different challenges faced by each country (International Food Policy Research Institute, 2021) and the different opportunities to address them.

The information gathered at country level can help us identify where the problem is most acute.

In order to address our first research question about the extent to which research on hunger focuses on countries afflicted by hunger, we calculated the aggregate number of country mentions for each category.

To address our second research question about the country affiliation of hunger research scholars, we calculated the aggregate number of papers in each GHI country category. We presented the country mentions and the country affiliations as absolute numbers and normalised for combined population of the country category.

To answer the first two research questions more fully, we looked more deeply at the non-overlapping papers. For the mentions-only papers and the affiliations-only papers, we performed bibliometric analyses and we conducted manual examination of random samples of the non-overlapping papers.

The mentions-only papers mention a country but do not feature any authors in the country mentioned. In order to discover where the authors are based, we quantified the number of country mentions for the country with most publications in the SDG 2 dataset from each GHI category.

We then manually examined the full text of 100 mentions-only papers (20 selected at random from each GHI category) to gain insight into the possible reasons that no locally affiliated authors were listed on the paper.

The affiliations-only papers do not mention the countries that the authors are affiliated to. We aim to find out whether these papers mention any country at all by comparing the affiliations-only papers with the SDG 2 papers that include no country mentions.

We then manually examined the full text of 100 affiliations-only papers (20 selected at random from each GHI category) to look into reasons why authors have not mentioned the country in which they are affiliated.

For the sample analyses, we grouped countries by GHI categories and assigned each paper to the most severe category it could belong. For example, a publication that mentioned a country from the low hunger category and another country from the serious category was only counted in the serious category. This way, we avoided papers being counted in multiple categories.

To address our third question about the equality of research collaborations, we determined the author position for each paper and aggregated them for each GHI country category. We then presented the share of first author country affiliations, last author affiliations, and overall country affiliations from each of the country categories.

# Results

## Geographical focus (country mentions)

The Dimensions SDG 2 database contains 59,778 papers published between 2016 and 2021 that are related to SDG 2 – Zero hunger. In one third of these papers, the authors mention at least one country in the title or abstract. In the context of our first research question, we note that 26% of the papers mention countries listed in the Global Hunger Index 2021 report. Within the same Dimensions SDG 2 dataset, 31,769 (53%) of the 59,778 publications featured author affiliations in countries listed in the categories of the GHI 2021 report. The remaining 47% of hunger research papers did not feature any authors from countries assessed for hunger by the GHI. The number of countries with their combined populations and number of author country affiliations are shown in Table 2. There is only one country in the extremely alarming category (Somalia), and we have therefore combined the two most severe categories as 'alarming / extremely alarming'. We also show the number and combined population of countries not assessed for the GHI. The 'not assessed' category includes many of the world's wealthy and economically developed nations. For a full list of country categorisation, see Appendix A.

Table 2. GHI country categories by population, country mentions and affiliations.

| GHI Hunger category | # Countries | Combined population | Mentions | Mentions per pop (m) | Affiliations | Affiliations per pop (m) |
|---|---|---|---|---|---|---|
| Alarming / extremely alarming | 10 | 232 | 434 | 1.87 | 288 | 1.24 |
| Serious | 37 | 2610 | 7281 | 2.79 | 12873 | 4.93 |
| Moderate | 31 | 1214 | 4475 | 3.69 | 7870 | 6.48 |
| Low | 50 | 2624 | 4776 | 1.82 | 13973 | 5.33 |
| Not assessed | 105 | 1235 | 4938 | 4.00 | 35834 | 29.02 |

The number of papers that mention countries in the most severe hunger categories is by far the lowest. The Mentions per pop (m) column presents the number of papers that mention countries within a category normalised for the combined population of those countries (Figure 2).

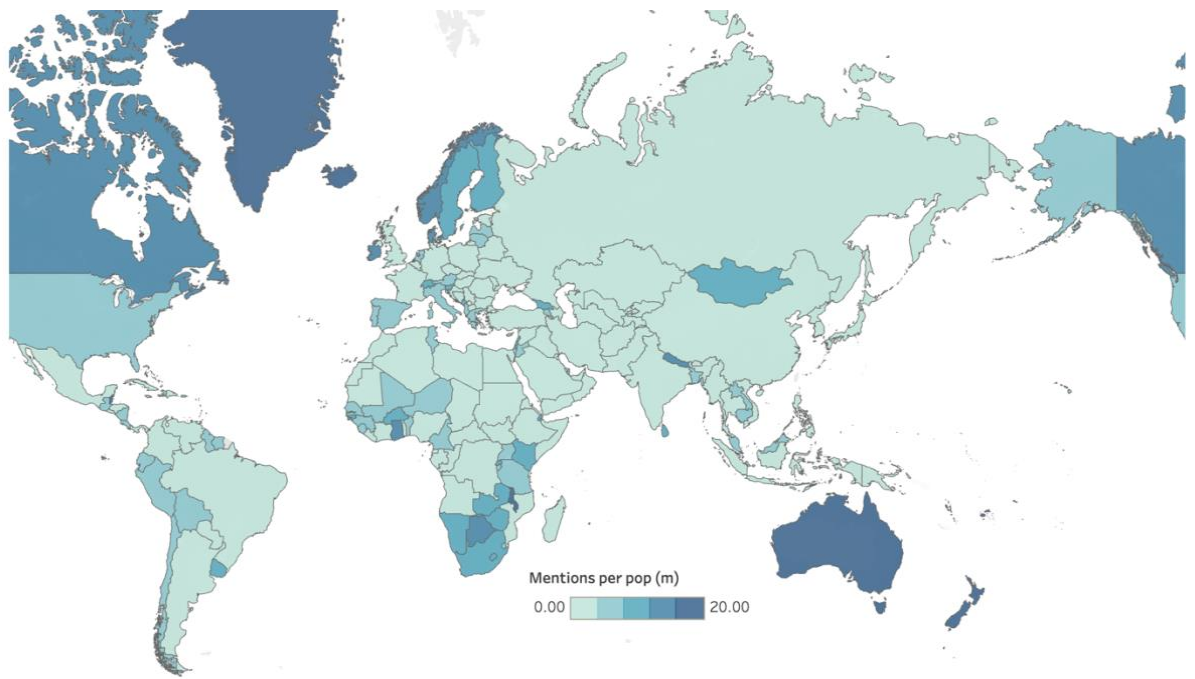

Figure 2. Mentions per million population by country

Within the GHI country categories, we see an interesting pattern. First, the countries in the low category have the fewest mentions per population. The moderate countries have more than twice the number of mentions per population as the low countries. However, as the severity of hunger increases from moderate, to serious, and then to alarming / extremely alarming, the number of mentions per population declines again. The frequency of mentions per population is higher for countries not assessed by the GHI (4.00) than for any of the categories that are assessed.

Geographical location (author affiliations)

Using author country affiliations, we assessed the geographical location of scholars involved in publishing hunger research (Table 2). Due to the variation in population between the countries in the GHI categories, the population normalised country affiliations are presented on a world map in Figure 3.

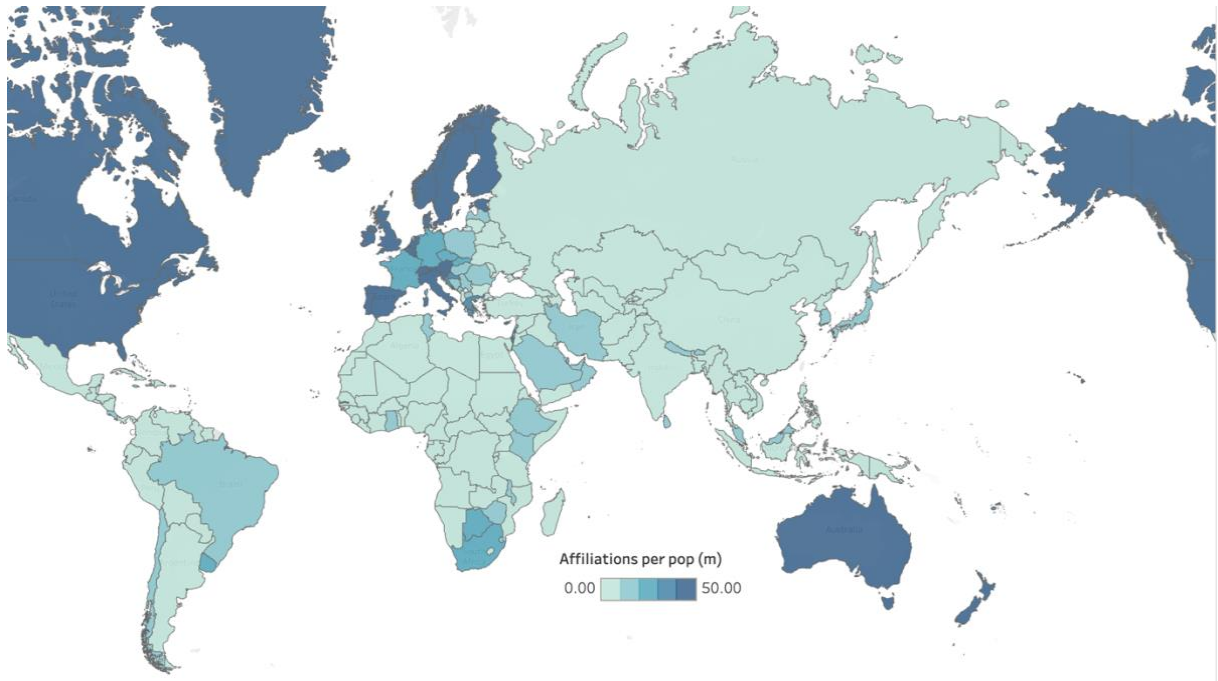

Figure 3. Author affiliations per million population by country.

Similar to the geographical focus, we found a downward trend in geographical author affiliations as the severity of hunger increased. Within the GHI categories, the affiliations per population peaks in the moderate category and then declines markedly as the severity of hunger increases. Indeed, researchers from countries most afflicted by hunger are twenty times less likely to publish research papers on hunger than those in wealthy countries.

Relationship between regional focus and regional authorship

The relationship between country mentions and country affiliations is shown by GHI category in Figure 4.

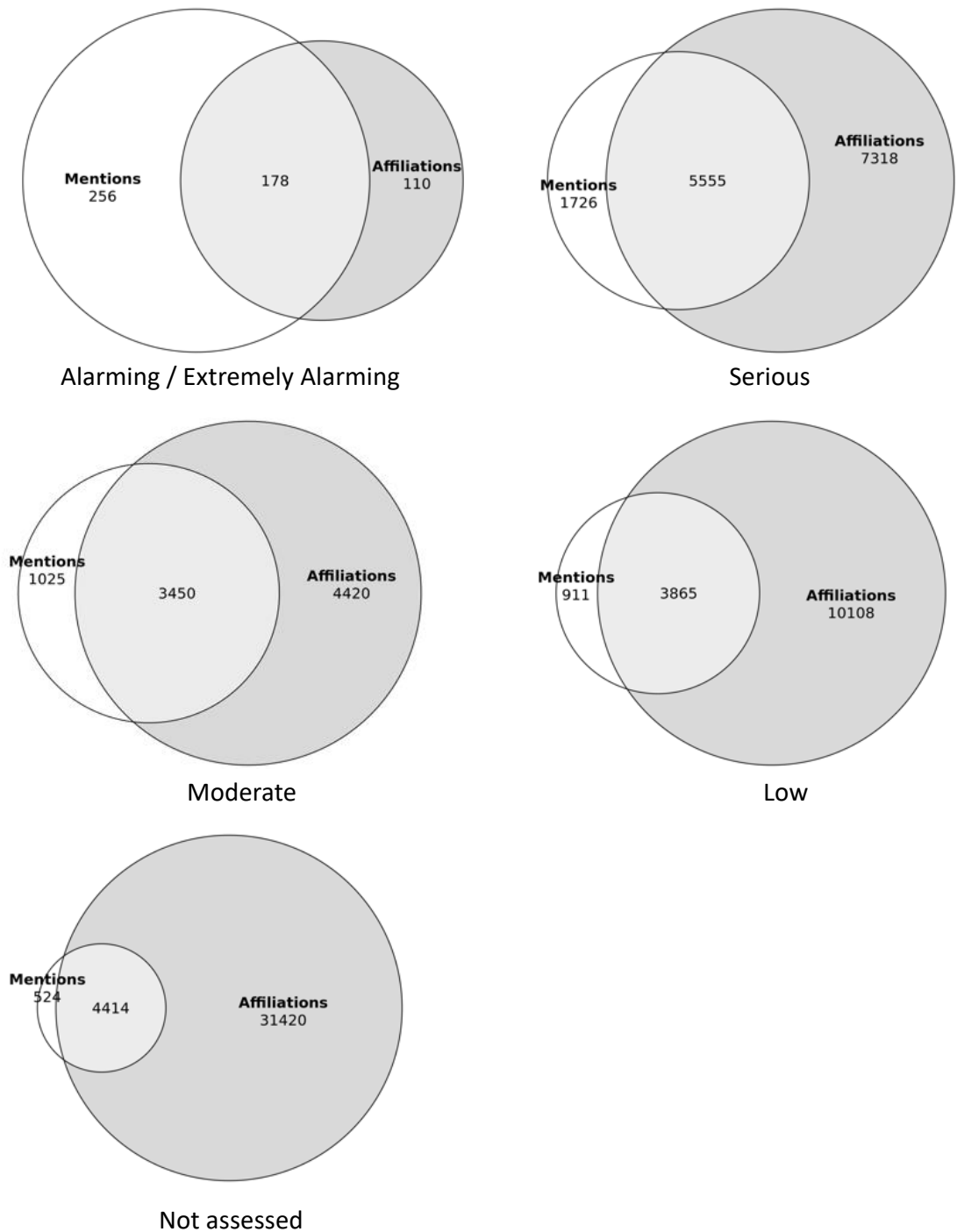

Figure 4. Mentions and affiliations by GHI categories.

For each GHI category, the circle on the left represents the number of papers which mention at least one country in that category. If a paper mentions the same country more than once or more than one country in the same category, it still only counts as one paper for that category. However, it the paper mentions one or more papers in two categories, then it counts as one paper for each category. The circle on the right follows the same rules but refers to author's country affiliation.

The share of mentions-only increased in line with the severity of hunger, the greatest share seen in the alarming or extremely alarming categories.

In Figure 5, we present the share of papers that mention countries from each GHI category that also feature at least one author from the same category (blue line). The orange line shows the share of author affiliations by GHI category that also focus on countries in the same category. For instance, the blue line shows that 41% of the papers that mention alarming or extremely alarming countries also feature at least one author affiliation from these categories. This is the lowest share and as the hunger severity decreases, the share of country mentions that also feature authors from the same category increases.

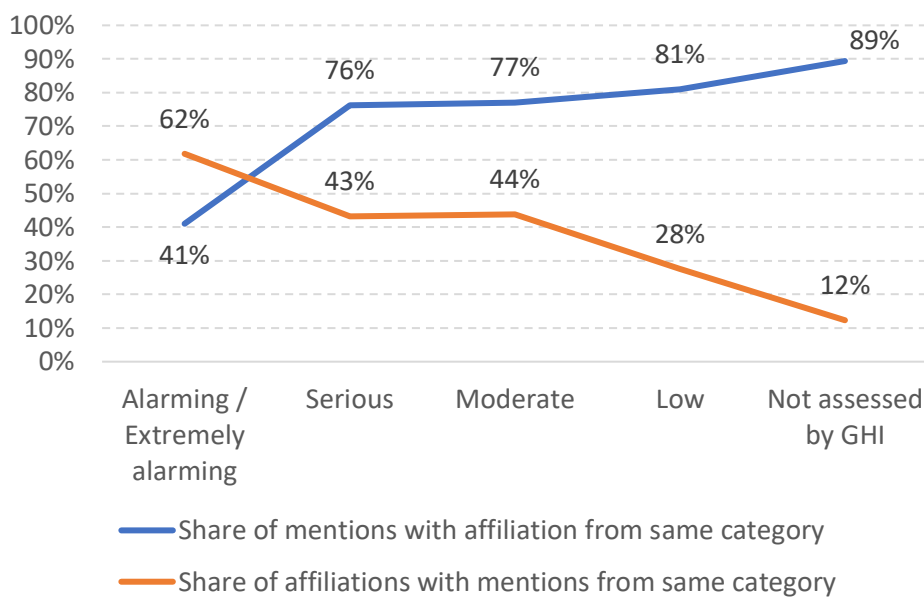

Figure 5. Share of mentions and affiliations by country category

The share of author affiliations by GHI category that also mention countries from the same category decreases as the severity of hunger decreases (orange line). Of the papers with authors from countries with the most severe hunger problems, 62% also mention countries from the same category. This is the case for only 12% of the papers with authors from countries in the not assessed category.

### Lack of local authors (mentions-only papers)

The mentions-only papers do not feature authors affiliated to the countries mentioned in the article. We would like to know where they are located. In 4 we show the most frequently occurring country affiliations on the mentions-only papers for each GHI category.

Table 3. Location of authors on mentions-only papers.

| Democratic Republic of the Congo mentions (Alarming) | Papers | % Share | India (Serious) | Papers | % Share | Indonesia (Moderate) | Papers | % Share |
|---|---|---|---|---|---|---|---|---|
| United States | 25 | 24% | United States | 195 | 8% | United States | 23 | 3% |
| United Kingdom | 7 | 7% | United Kingdom | 74 | 3% | India | 15 | 2% |
| India | 3 | 3% | Australia | 41 | 2% | Australia | 13 | 2% |
| South Africa | 3 | 3% | China | 31 | 1% | Japan | 12 | 2% |
| Australia | 3 | 3% | Germany | 30 | 1% | United Kingdom | 11 | 1% |
| Belgium | 3 | 3% | Canada | 29 | 1% | Thailand | 7 | 1% |
| China | 3 | 3% | Switzerland | 17 | 1% | Germany | 6 | 1% |
| Canada | 2 | 2% | South Africa | 14 | 1% | South Africa | 6 | 1% |
| Denmark | 2 | 2% | Bangladesh | 14 | 1% | Switzerland | 6 | 1% |
| France | 2 | 2% | Japan | 13 | 1% | France | 5 | 1% |
| China (Low) | Papers | % Share | USA (Not assessed) | Papers | % Share | | | |
| United States | 83 | 5% | India | 48 | 3% | | | |
| India | 62 | 4% | United Kingdom | 44 | 2% | | | |
| United Kingdom | 40 | 2% | China | 41 | 2% | | | |
| Australia | 30 | 2% | Australia | 38 | 2% | | | |
| Germany | 23 | 1% | Canada | 30 | 2% | | | |
| Netherlands | 20 | 1% | Italy | 16 | 1% | | | |
| Japan | 16 | 1% | Iran | 14 | 1% | | | |
| Canada | 14 | 1% | Germany | 13 | 1% | | | |
| France | 12 | 1% | Russia | 12 | 1% | | | |
| Italy | 12 | 1% | Brazil | 11 | 1% | | | |

In all groups, many of the mentions-only papers were published by authors affiliated to countries with large populations and high publication output such as the United States, United Kingdom, China, and India. There were however, some prominent appearances from regional neighbours. In the extremely alarming and alarming category (Democratic Republic of the Congo), there were publications from regional neighbour South Africa, and from Belgium which has traditional ties to the DRC. In the serious category (India), there are publications from authors in neighbouring

Bangladesh, while in the moderate category (Indonesia) there are publications from Thailand and Japan.

We also present an analysis of 100 randomly sampled mentions-only papers (20 for each of the GHI categories) (Table 4). We examined the full text of each paper in the sample to discover the reason for the lack of country affiliation.

Table 4. Mentions-only publication level analysis

|  | Alarming / Extremely alarming | Serious | Moderate | Low | Not assessed |
|---|---|---|---|---|---|
| International researchers focusing on a different region | 13 | 15 | 17 | 12 | 3 |
| Researchers affiliated to neighbouring country | 3 | 2 | 3 | 0 | 1 |
| Passing mention of country but no topical focus | 4 | 3 | 0 | 8 | 16 |

Sample data is made available in Zenodo (Purnell, 2023)

We found that while authors may not be in the same country as that mentioned in the title or abstract, it doesn't mean they are very remote. We found several papers that featured authors in neighbouring countries to those mentioned. In one case (10.1186/s12889-020-08657-x), an international group sought to improve the criteria for community-based treatment of malnutrition in South Sudan. Some of the authors were based in Kenya, which shares a border with South Sudan.

Some studies were not focused on countries that were mentioned in the title or abstract of the paper. This was caused by passing mention of a country. For instance, in the abstract of a study of a rice production system used throughout India (DOI: 10.1007/978-981-10-3692-7), the authors mention the method originated in Madagascar, and although Madagascar is not mentioned again, the paper has been classified as focusing on a country in the alarming category but without local authors. On these papers the country name was mentioned in the title or abstract but no author affiliations were present from that country. Lack of country mentions (affiliations-only papers)

The affiliations-only papers make no mention of the countries in with the authors are affiliated. We wanted to know whether these papers mentioned any country at all. In Table 5 we show the number of affiliations-only papers for the country with the highest number of hunger-related research papers for each of the GHI country categories. We also show the number and share of affiliations-only papers that did not mention any country at all.

Table 5. Papers with no country mention.

| Country name | GHI category | Affiliations-only | No country mentioned | Share with no country mentioned |
|---|---|---|---|---|
| Democratic Republic of the Congo | Alarming / Extremely alarming | 37 | 24 | 65% |
| India | Serious | 5119 | 4915 | 96% |
| Indonesia | Moderate | 995 | 952 | 96% |
| China | Low | 3936 | 3615 | 92% |
| United States | Not assessed | 13058 | 9771 | 75% |

Sample data is made available in Zenodo (Purnell, 2023)

In all GHI categories, a high share of affiliation-only papers did not mention any country name at all. the lowest share was in the Democratic Republic of the Congo whose authors mentioned no country in 65% of the affiliations-only papers.

We also conducted an analysis of 100 randomly sampled affiliations-only papers (20 for each of the GHI categories) (Table 6). We examined the full text of each paper in the sample to determine the reason for the lack of country mention.

Table 6. Reasons for not mentioning country of author affiliation.

|  | Alarming / Extremely alarming | Serious | Moderate | Low | Not assessed |
|---|---|---|---|---|---|
| Study focuses on other region, or has no regional focus | 13 | 17 | 13 | 9 | 17 |
| Local region is mentioned but not the country name | 4 | 3 | 5 | 4 | 2 |
| Local study but region not mentioned | 3 | 0 | 2 | 7 | 1 |

One common reason for not mentioning the country where the authors are based is that the article has no regional focus. For instance, we found several papers with author affiliations in the serious category were published by authors in India and focused on methods for improving agricultural methods and had no regional focus.

A second group of affiliations-only papers were indeed studies that focused on a specific country, but the authors had not mentioned the country at all in the title or abstract of the paper. One example (10.1155/2019/4740825) is a clinical study with all the authors based in the Democratic Republic of the Congo and thanking a local hospital in the acknowledgements. The study is clearly focused on the Democratic Republic of the Congo but the authors have not mentioned the country in the title or abstract of the paper.

In a third set of affiliations-only papers, the authors do mention the local site of the study but not necessarily the country name. We found evidence of authors using city names e.g., Royapettah (10.18203/2349-3291.ijcp20174153), islands e.g., Zanzibar (10.1093/cid/cix500), or simply referring to 'our nation' (10.1007/978-981-16-6124-2_1).

### Partnerships

To address our third research question about the equality of international partnerships, we analysed the author position in more detail. For 55,422 (93%) of the SDG 2 papers, we could identify the country affiliation of the first and last authors. We found that the last author has an affiliation in a country listed in the GHI 2021 categories on 40% of the SDG 2 papers.

Considering all SDG 2 papers, only 0.35% of author affiliations were from countries in the alarming or extremely alarming GHI country categories (

Figure 6). The figure was even lower for authors in first (0.23%) and last author (0.17%) positions. Meanwhile, the combined population of the countries in the alarming and extremely alarming countries comprises 2.9% of the world's population. Authors from these countries are therefore underrepresented in scholarly research on hunger and especially underrepresented in lead author positions.

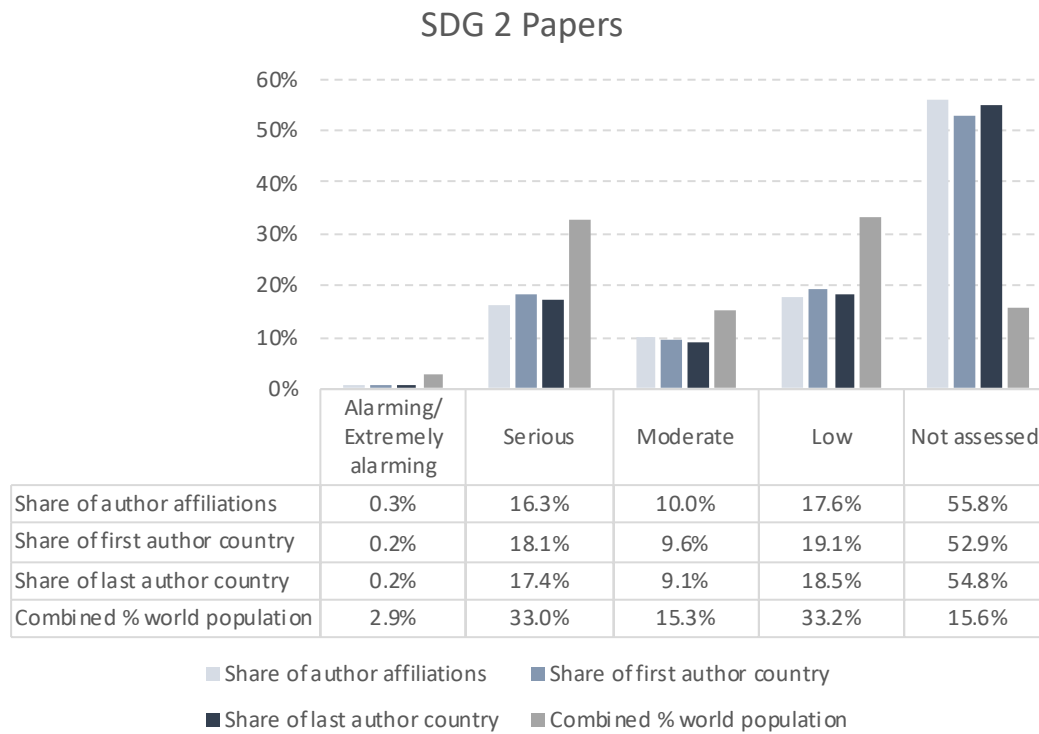

Figure 6. Author position - All SDG 2 papers.

Authors from countries in the serious, moderate, and low GHI categories are also underrepresented when compared with their share of the world's population albeit to a lesser extent. The share of lead author positions is not very different from overall authorship. In those countries not assessed by the GHI, which includes relatively wealthy countries, the pattern is reversed with authors overrepresented compared with their combined share of the world's population.

# Discussion

We used country mentions and author country affiliations to study the geographical diversity of researchers, using scholarly papers related to SDG 2: Zero hunger as a case study. For ease of comparison, we used the country categories as described in the 2021 Global Hunger Index report which are based on the severity of hunger.

About one third of hunger research papers mention at least one country in their title or abstract, although this trend diminished when mentioning the country categories most severely affected by hunger. This was a counterintuitive result, and we consider possible explanations. First, there are only ten countries in the alarming and extremely alarming categories compared with 118 in the less severe categories and 106 not assessed countries. That means countries in less severe categories comprise more country names that are available to be mentioned. Similarly, the smaller relative population in the alarming categories means there are fewer local authors to focus on their own country than in the less affected categories.

We cannot be sure of authors' motives when they decide to name a country in their paper. There is evidence that authors from wealthy, developed countries with established research infrastructure, especially the United States, are less likely to mention the country names even in locally based studies (Castro Torres & Alburez-Gutierrez, 2022; Kahalon et al., 2021). Consequently, there are potentially studies with a country focus that do not mention the country of focus, and that those papers missed are predominantly in the lower GHI country categories. That means the real tendency of decreasing topical focus in countries less affected by hunger may not be as sharp as our results show.

We used author affiliations to determine their geographical location and defined local authors as those whose country affiliation matched the name of the country mentioned in the paper. We found the presence of local authors declined as the severity of hunger in the mentioned country increased, especially on papers that mentioned countries in the alarming and extremely alarming categories which are the most severe. We suggest potential reasons for the lack of local authors on research papers that focus on the most severely affected countries.

Countries most severely affected by hunger are also among the world's poorest countries and poverty can be a barrier to young people entering higher education. The related impact of long-term poverty, especially in countries experiencing war or civil conflict is limited the development of a research infrastructure in some countries categorised as suffering from alarming levels of hunger. Less developed research infrastructure is therefore proposed a possible explanation for the lack of local academics and associated research publications in the most severe GHI categories. Meanwhile, academic publishing has flourished in some countries with large populations in the less severe hunger categories such as India (serious) and Indonesia (moderate). The prolific scholarly output in these countries has contributed to relatively large numbers of country mentions in hunger research.

That researchers from countries not assessed by the GHI publish the majority of SDG 2 papers is not surprising. The GHI doesn't assess developed, wealthy countries and these countries traditionally publish the most research. In 2015, the UN called for the whole world, not just those living in affected areas, to respond to a series of challenges. The contribution of academics in developed nations could be interpreted as an encouraging sign that the global research community is engaged in addressing the grand challenges of our time. It may also be symptomatic of a sampling bias caused by overrepresentation of country affiliations from high income countries (Bylund et al., 2023).

In all the GHI categories, our results showed in most cases that authors either mention the name of their own country or they do not mention any country at all. Papers that do not mention any country may have no regional focus or the authors decided not to mention the name of the country in which the study was conducted.

The appearance of local contributors mentioned in the acknowledgements section of the article but who did not appear as co-authors might also partially explain the absence of local authors. In some papers, collaborators from the country of focus were acknowledged and thanked in the publication even if they did not feature as authors of the study. Acknowledgements form part of the reward triangle bestowed by researchers on those who have helped or significantly influenced academic publications (Costas & van Leeuwen, 2012; Cronin & Weaver, 1995). The difference between a contribution that deserves 'only' acknowledgement, and one worthy of co-authorship is a key distinction. The observation of acknowledgements in our analysis of sample papers warrants further examination to see whether this phenomenon has any relation with the GHI country categories.

### Author position and partnership dynamics

Author position in the context of partnerships between visiting and local researchers has come under scrutiny by the academic publishing community (e.g., Morton et al., 2022; Nature, 2022; PLOS, 2021; van Groenigen & Stoof, 2020). This study contributes to the discussion by demonstrating the relationship between author position and affiliation from both geographical focus and geographical location perspectives.

We specifically identified the country affiliation for authors who appeared in first or last author position. First, and especially last author position has been used as a proxy for lead authorship in several studies. We found that authors from countries in the most severe GHI categories were underrepresented in lead author positions. However, in all the other categories the share of lead author positions was close to the overall share of authors. In the serious and low GHI categories, local authors were slightly overrepresented compared with all author positions. However, where countries from GHI categories were the focus of research, authors from the same category featured far more frequently and were even overrepresented in lead author positions. This result is suggestive of greater participation of local academics in studies with a regional focus.

Our study builds on earlier work conducted in health fields that reported underrepresentation of local researchers in lead author positions (Hedt-Gauthier et al., 2019; Mbaye et al., 2019). Our results support the idea that unequal partnerships may exist in international collaborative research, but only in countries severely affected by hunger. In all other categories, our author position analysis showed that the share of local researchers in lead author positions was similar to the overall share of local authors. We therefore encourage further discussion on the partnership dynamics between local academics and international collaborating partners.

These findings could be interpreted as showing that international academics from wealthy nations more frequently lead studies on the most severely affected regions than local authors. Also the idea that local contributors might play a limited role such as data collection but not authorship on large international studies (Asiamah et al., 2021).

### Study limitations

This study was the first to examine the geodiversity of research using SDG 2 as a case study. However, we acknowledge the study has a number of limitations that could be used to identify areas for follow on studies.

First, the GHI country categories are based on a composite indicator. The cut-off scores that assign countries to one category or another are necessarily arbitrary but mean that there might be greater differences between countries at the extremities of a category than between countries separated by the cut-off score. We also hypothesise that conditions may vary within a country, such that different regions of a country would meet the conditions for different GHI country categories. We have presented evidence of community-based collaboration across international borders between countries that have landed in different GHI categories. The GHI categories are therefore not infallible in their indication of severity of hunger. We could have used nation states grouped by geographical or political regions instead, but those groupings would of course be subject to the same limitations as the GHI categories.

The definition of the body of research papers on hunger research is not unequivocal. We used the Dimensions database because of its inclusive coverage (Hook et al., 2018; Huang et al., 2020; Visser et al., 2021) and AI algorithm used to tag papers related to the different SDGs. However, competing bibliometric databases have used slightly different approaches to identifying SDG research which make huge differences to the papers retrieved. Comparisons have shown very limited overlap between databases in specific SDG searches (Armitage, Lorenz, & Mikki, 2020; Purnell, 2022). Using alternative publication data sources for SDG related research studies could have therefore produced different results in our study. Indeed, the varying search techniques and use of AI algorithms between publication databases is a limitation now applicable to every bibliometric study that uses SDG related research.

While Dimensions coverage is broad, it is not comprehensive and relies mainly on Crossref and PubMed as it sources. For publications to be in Crossref, they need to register a digital object identifier (DOI). This requires some expertise and a financial arrangement where publishers pay $1 US for each paper assigned a DOI. While these arrangements may be easily achievable in some regions, in other less economically developed areas of the world they might pose a barrier to publishing. In these cases, papers are more likely to appear in local university presses, not be assigned a DOI, and therefore not be indexed by Dimensions. Unfortunately, poorer countries appear in the most severe GHI categories and are at the highest risk of lower database coverage (Giménez-Toledo, Mañana-Rodríguez, & Sivertsen, 2017). As a consequence, there might be proportionately more research papers on hunger published by authors affiliated to the most affected countries not indexed in Dimensions and therefore missing from our study. Crossref has recently announced an initiative to address this obstacle (Collins, 2022).

We have used mention of a country name in the title or abstract of an article as an indicator that the study focuses on that country to a certain extent. However, our manual examination of sample papers uncovered evidence of countries mentioned in passing and that do not represent the geographical focus of the study. For instance, in a study of a rice production system used throughout India (DOI: 10.1007/978-981-10-3692-7), the authors mention Madagascar in the abstract as the country where the method originated. In this case, the mention of Madagascar does not denote topical focus of the article.

This finding shows that a single appearance of a country name should not alone be accepted as a reliable indicator of geographical focus. Future studies could look for ways to improve knowledge surrounding the use of country mentions as indicators of focus such as interpreting the country name in context. The passing mentions in our study only mention a country once and consequently multiple mentions of a country name in an article might strengthen the indication of focus.

Conversely, we have assumed that the absence of country name in the article title or abstract means the study was either focused somewhere else or had no regional focus at all. This assumption might not always hold true. Scholarly papers by authors in wealthy countries with long establish research infrastructure often do not mention the name of the country even in locally focused studies (Castro Torres & Alburez-Gutierrez, 2022; Kahalon et al., 2021). Some authors refer to regions of the world, e.g., Sub-Saharan Africa rather than countries, or to smaller units of countries like cities, islands, or regions. These examples serve as further evidence to support research into the use of country mentions as indicators of geographical focus.

## Conclusion

This is the first large-scale bibliometric study on the geodiversity of research that used country mentions as an indicator of topic focus and author affiliations to identify geographical location. In light of the urgent and growing problem presented by hunger, we used the body of research papers related to SDG 2: Zero Hunger in the Dimensions database as a case study, and the Global Hunger Index (GHI) country categories for ease of comparison.

We found that hunger research papers focused less on the countries in the most severe GHI country categories (extremely alarming or alarming) although that may be partially explained by the comparatively low aggregate population of those countries. On the majority of papers that mentioned countries in the alarming and extremely alarming GHI country categories, the were no authors from the country mentioned. Instead, the majority of authors on the mentions-only papers were from relatively populous and wealthy countries. However, we conducted a manual examination of random samples of the mentions-only papers and found that some authors were based in neighbouring countries.

There were also methodological reasons for the low rate of mentions of countries most severely affected by hunger. There are fewer countries in the severe GHI countries to mention and the aggregate population is relatively small. Researchers based in wealthy countries are also less likely to mention the geographical focus of their study than authors in the severe GHI country categories. Use of country mention as an indicator of geographical focus might therefore not be uniform across the countries in this paper and future studies could investigate further.

Our study showed declining participation of local authors affiliated to institutions in the most severe GHI country categories. Author affiliations per population in the extremely alarming and alarming categories was one-twentieth of that in the wealthy countries not assessed by the GHI. Authors based in the most severe GHI country categories mentioned their country in most cases but not in the less severe country categories. We found in all country categories, that most affiliations-only papers did not contain mention of any country at all. Either the authors chose not to mention the country of the study, or the study had no regional focus.

Our examination of a random sample of affiliations-only papers revealed that in some studies with a localised geographical focus simply omitted to mention the site of the study, while others mentioned the city, island, or region name but not the country name. Again, this shows the shortcomings of using country names as a reliable method of capturing the whereabouts of all authors. To improve recall, the method could be extended to cover additional geographical terms. Future studies would then more accurately determine the share of studies that have no regional focus.

We observe that many of the countries in the most alarming categories have also faced civil conflict, famine, and other causes of long-term instability that may limit the development of research infrastructure and consequent publication output.

The underrepresentation of academics from countries most affected by hunger in last author position is of concern in the context of equitable and fair international research collaboration. Our manual analyses of sample papers in the extremely alarming and alarming categories showed that sometimes local contributors are mentioned in the acknowledgements section of the paper, rather than being included as co-authors. We encourage further analysis on the criteria that warrant co-authorship and whether it is justly applied across all countries. The trend was not repeated in the other categories and our findings therefore did not support the notion of widespread unequal publication practice. Indeed, the examples in our manual analyses did not seem to resonate with the 'helicopter' or 'parachute' research practices described in recent literature (e.g., Heymann et al., 2016; Minasny & Fiantis, 2018; North et al., 2020).

Manual examination of the papers revealed that other factors were at play. For instance, the assumption that non-local authors are based in safe, wealthy countries is often wrong. Many are in fact located just across an international border or based in a country in an even more severe hunger category than their co-authors. We also found that some papers that mentioned a country didn't really focus on it and the lack of authors from that country could not therefore be interpreted as evidence of questionable research practice.

In our view, reports of helicopter research that call into question researchers' motives require clear definitions and methods that include a level of qualitative assessment including at the very least, manual examination of the publication. While some high-profile examples of questionable research ethics have been published, we urge caution when extrapolating superficial metrics such as author affiliation or author position to ensure they don't lead to unfounded conclusions (e.g., Nature, 2022).

## Acknowledgements

Ton van Raan and Ludo Waltman for expert advice and guidance throughout the study. Marja Spierenburg, Rodrigo Costas, Ismael Rafols, and the two anonymous reviewers for feedback on earlier versions of the paper.

## Declarations

Competing interests

The authors have no relevant financial or non-financial interests to disclose.

## Data availability

The sample records manually examined summarised in tables 3 and 4 are available in Zenodo (Purnell, 2023)

*Studies*, *1*(2), 445–478. https://doi.org/10.1162/qss_a_00031

International Food Policy Research Institute. (2021). IFPRI Country Programs. Retrieved November 23, 2021, from https://www.ifpri.org/topic/ifpri-country-programs

IUCN–UNEP–WWF. (1980). *World Conservation Strategy: Living Resource Conservation for Sustainable Development.* Retrieved from https://portals.iucn.org/library/efiles/documents/wcs-004.pdf

Kahalon, R., Klein, V., Ksenofontov, I., Ullrich, J., & Wright, S. C. (2021). Mentioning the Sample's Country in the Article's Title Leads to Bias in Research Evaluation. *Social Psychological and Personality Science*, *13*(2), 352–361. https://doi.org/10.1177/19485506211024036

Kalleberg, A. L. (2009). Precarious work, insecure workers: Employment relations in transition. *American Sociological Review*, *74*(1), 1–22. https://doi.org/10.1177/000312240907400101

Kates, R. W. (2011). *From the Unity of Nature to Sustainability Science: Ideas and Practice." CID Working Paper* (No. 218). Retrieved from https://dash.harvard.edu/bitstream/handle/1/37366242/218_Unityof Nature to Sustainabiltiy.pdf

Kates, R. W., Clark, W. C., Corell, R., Hall, J. M., Jaeger, C. C., Lowe, I., … Svedin, U. (2001). Sustainability Science. *Science*, *292*(5517), 641–642. https://doi.org/10.1126/science.1059386

Lai, B., & Thyne, C. (2007). The Effect of Civil War on Education, 1980—97. *Journal of Peace Research*, *44*(3), 277–292. https://doi.org/10.1177/0022343307076631

Leaning, J., & Guha-Sapir, D. (2013). Natural Disasters, Armed Conflict, and Public Health. *New England Journal of Medicine*, *369*(19), 1836–1842. https://doi.org/10.1056/NEJMra1109877

Mammides, C., Goodale, U. M., Corlett, R. T., Chen, J., Bawa, K. S., Hariya, H., … Goodale, E. (2016). Increasing geographic diversity in the international
32